\documentclass[journal]{IEEEtran}
\usepackage[cmex10]{amsmath}
\usepackage{amssymb}
\usepackage{eucal}
\usepackage{array}
\usepackage{eqparbox}
\usepackage{color}
\usepackage[normalem]{ulem}
\newcommand{\be}{\begin{equation}}
\newcommand{\ee}{\end{equation}}
\newcommand{\bd}{\begin{displaymath}}
\newcommand{\ed}{\end{displaymath}}
\newcommand{\ba}{\begin{eqnarray}}
\newcommand{\ea}{\end{eqnarray}}

\def\smooth {{\mathbb S}}

\def\d{\delta}

\def\v12{(v-w)}

\def\({\left(}
\def\){\right)}

\def\bgr#1\egr{{\allowdisplaybreaks\begin{gather}#1\end{gather}}}
\def\bma#1\ema{{\allowdisplaybreaks\begin{align}#1\end{align}}}

\def\oplem#1{\begin{lemma}\, {\rm #1}\, \it }
\def\cllem{\end{lemma}\rm \par }
\def\opthm#1{\begin{theorem}\, {\rm #1}\, \it }
\def\clthm{\end{theorem}\rm \par }

\newcommand{\R}{\mathbb R}%\def\R{{I \!\! R}}

\newcommand{\fer}[1]{(\ref{#1})}
\newcommand{\bq}{\begin{equation}}
\newcommand{\eq}{\end{equation}}
\def\bqa{\begin{eqnarray}}
\def\eqa{\end{eqnarray}}
\def\bd{\begin{displaymath}}
\def\ed{\end{displaymath}}

\newtheorem{thm}{Theorem}

\renewcommand{\d}{\mathrm d}
\newcommand{\rme}{\mathrm e}
\newcommand{\rmD}{\mathrm D}
\newcommand{\sffp}{q}
\newcommand{\sfep}{\mathsf e_p}
\newcommand{\BBB}{}

\newcommand{\nc}{\normalcolor}
\newcommand{\cH}{\mathcal H}
\newcommand{\cI}{\mathcal I}
\newcommand{\rH}{\mathrm H}
\newcommand{\rE}{\mathrm E}
\newcommand{\rD}{\mathrm D}
\newcommand{\rF}{\mathrm F}
\newcommand{\rI}{\mathrm I}
\newcommand{\cN}{\mathcal N}
\newcommand{\rN}{\mathrm N}
\begin{document}
\title{The concavity of R\'enyi entropy power}
\author{Giuseppe~Savar\'e
        and~Giuseppe~Toscani% <-this % stops a space
\thanks{G. Savar\'e and G. Toscani are with the Department
of Mathematics, University of Pavia, 27100 Pavia, Italy (e-mail:
giuseppe.savare@unipv.it; giuseppe.toscani@unipv.it)
 (see http://www-dimat.unipv.it).}% <-this % stops a space
\thanks{ The authors acknowledge support by MIUR project
``Optimal mass transportation, geometrical and functional
inequalities with applications''.}% <-this % stops a space
\thanks{Manuscript received ; revised .}}

%\markboth{$\null$}
%{Shell \MakeLowercase{\textit{G. Savar\'e et al.}}:$\null$% IEEE
%TRANSACTIONS ON INFORMATION THEORY}

\maketitle

\begin{abstract}
%\boldmath
We associate to the $p$-th R\'enyi entropy a definition of entropy power,
which is the natural extension of Shannon's entropy power
\BBB and exhibits a nice behaviour along solutions to
the $p$-nonlinear heat equation in $\R^n$.
We show that the R\'enyi entropy power of general probability
densities solving such equations is always
a concave function of time,
whereas it has a linear behaviour in correspondence to the
Barenblatt source-type
solutions. \nc
% it is
% We then shown that the $p$-th R\'enyi entropy power of a
% probability density which solves the nonlinear diffusion of order $p$, is
% a concave function of time.
This result extends Costa's concavity
inequality for Shannon's entropy power to R\'enyi entropies.
\end{abstract}


\section{Introduction}
\IEEEPARstart T{} he $p$-th R\'enyi entropy of a random variable $X$ with
density $f$ in $\R^n$ is defined by (see, e.g. Cover and Thomas \cite{CT}
and Gardner \cite{Gar})
 \be\label{ren}
 \BBB \cH_p(X) \nc
% \footnote{\BBB Ho cambiato il simbolo per distinguere
%   l'entropia sulla random variable e quella definita sulle densit\`a:
%   si potrebbe scrivere tutto in termini delle densit\`a?}
= \rH_p(f) := \frac 1{1-p} \log \int_{\R^n} f^p(x) \, \d x,
 \ee
for $0<p< +\infty$, $p \not= 1$.

Whenever \BBB
$p > 1-2/n,\nc$
% \footnote{\BBB si legge meglio come mai
% il coefficiente della (2) \`e positivo}
we
consider the positive coefficients
\begin{equation}
  \label{eq:3}
  \mu:=2+n(p-1),\quad
  \nu%=\mu_{p,n}
  :=\frac{\mu}n=\frac 2n + (p-1),
\end{equation}
\nc
and we associate to the $p$-th
R\'enyi entropy
\BBB the \nc
entropy power
\BBB (that we call \emph{$p$-th Renyi entropy power} in
the following)
%\footnote{oppure vuoi evitare che venga implicitamente attribuita a Renyi...?}
\nc given by
 \be\label{pow-ren}
 \cN_p(X) \BBB =\rN_p(f)\nc:= \exp\big( \nu\,%\frac 2n +
                                %(p-1)\right)
    \cH_p(X) \big). \quad
  %\right\}, \quad
%  \mu:=\frac 2n + (p-1).
 \ee
% \sout{According to the lower bound on $p$,  the random variable $X$ possesses an
% entropy power if and only if the coefficient $\mu$ in front of $\cH_p(X)$ in the
% exponential is positive.}
% \footnote{\BBB Se la frase non ha qualche senso pi\'u profondo,
%   lascerei l'osservazione implicita nella definizione di $\mu$,.%  oppure \`e
% % solo un'osservazione? Altrimenti direi ``Notice that
% % the coefficient in front of $\cH_p(X)$ in the
% % exponential is nonnegative thanks to the lower bound
% % on $p$.''
% }
%
The R\'enyi entropy for $p=1$ is defined as the limit of $\cH_p(X)$ as
$p \to 1$. It follows directly from definition \fer{ren} that
%\footnote{\BBB Mi sembra che ci voglia un $-$ davanti all'entropia, giusto?}
 \[
\cH_1(X) = \lim_{p\to 1}\cH_p(X) = \cH(X) =\BBB -\nc \int_{\R^n} f(x) \log f(x) \,
\d x.
 \]
Therefore, the Shannon's entropy can be identified with the R\'enyi
entropy of index $p=1$. In this case, the proposed R\'enyi entropy power
of index $p=1$, given by \fer{pow-ren}, coincides with Shannon's
entropy power
 \be\label{pow}
 \cN(X) = \rN(f):=
 \nc \exp\left\{ \frac 2n \cH(X) \right\}.
 \ee
In 1985 Costa \cite{Cos} proved that, if $u(\cdot,t)$,
\BBB $t>0$, are probability densities solving
\nc  % is the solution to
the heat equation
 \be\label{heat}
\frac{\partial}{\partial t} u = \Delta u,
 \ee
posed in the whole space $\R^n$, then
 \be\label{conc}
\frac{\d^2}{\d t^2}\rN(u(\cdot,t)) \le 0.
 \ee
Inequality \fer{conc} is referred to as the \emph{concavity of entropy
power} theorem (see Riul \cite{Riu} for an exhaustive list of references).
The original proof of Costa has been simplified years later by Dembo
\cite{De1, DCT} with an argument based on the Blachman--Stam inequality
\cite{Bla}. Next, a direct proof of \fer{conc} in a strengthened form,
with an exact error term, has been obtained by Villani \cite{Vil}. The
proof in \cite{Vil} highlights a strong connection between the concavity
of entropy power and some identities of Bakry and Emery \cite{BE},
established through the so-called $\Gamma_2$ calculus as part of their
famous work on logarithmic Sobolev inequalities and hyper-contractive
diffusions. These connections, together with various consequences of the
concavity of entropy power theorem, have been recently discussed in
\cite{Tos}.

In this paper we show that the \emph{concavity of entropy power} is a
property which is not restricted to Shannon entropy power \fer{pow} in
connection with the heat equation \fer{heat}, but it holds for the $p$-th
R\'enyi entropy power \fer{pow-ren}, if we put it in connection with the
solution to the nonlinear heat equation
 \be\label{nl}
\frac{\partial}{\partial t} u = \Delta u^p,
 \ee
posed in the whole space $\R^n$. The precise result is the following.
\begin{thm}
  Let \BBB $p>1-2/n$ and let $u(\cdot,t)$ be probability densities in
  $\R^n$ solving \eqref{nl} for $t>0$. \nc
  % \footnote{\BBB Ho eliminato il dato iniziale, tanto non entra
  % nella disuguaglianza, e non ci serve neppure che la soluzione sia
  % unica.}  Let $f(x)$ be a probability density in $\R^n$ such that,
  % for $p > (n-2)/n$, the nonlinear heat equation \fer{nl}, with $f$
  % as initial datum, has a unique solution $u(x,t)$.
  Then the \BBB $p$-th R\'enyi entropy power \nc
  % entropy power associated to the $p$-th R\'enyi entropy
  defined in \fer{pow-ren} satisfies \be\label{conc-p} \frac{\d^2}{\d
    t^2}\rN_p(u(\cdot,t)) \le 0 \quad  \text{i.e.\ } t\mapsto
  \rN_p(u(\cdot,t))\text{ is concave.}  \nc \ee
\end{thm}
Like in the Shannon's case, inequalities \fer{conc-p} lied to sharp
isoperimetric inequalities. It is remarkable that the range of the
parameter $p$
%\RRR\sout{in the nonlinear heat equations}
\nc for which we can introduce the
R\'enyi entropy power, coincides with the range for which there is mass
conservation for the solution of \fer{nl} \cite{BDV}.

The rest of this paper is devoted to the proof of \fer{conc-p}. Before
entering into technical details, we will however explain in some details
the physical reasons which in our opinion suggest to define the
\BBB $p$-th R\'enyi entropy power \nc
%entropy power associated to R\'enyi entropy
in the form \fer{pow-ren}. To this
aim, we will introduce in Section \ref{nld} some known facts about
nonlinear heat equations (cfr. the book by Vazquez \cite{Vaz}, which fully
treats the case $p>1$, and \cite{BDV} for the case $p < 1$). Among other
facts, this connection between R\'enyi entropies and the nonlinear heat
equations allows to recover in a simple way a related $p$-th Fisher
information recently considered in \cite{LYZ}, \cite{JV}.

\section{\BBB Self-similar solutions \nc
% \sout{Nonlinear heat equations}
and R\'enyi entropies}\label{nld}

The relationship between Shannon's entropy power and the solution to the
linear heat equation \fer{heat} can be fruitfully highlighted owing to the
\BBB fundamental \nc
solution, representing heat release from a point
source
(here the origin $x=0$ without loss of generality). The source-type solution of
unit mass at time $t>0$ is represented by the Gaussian density
 \be\label{sol-h}
M(x,t) := \frac 1{(4\pi t)^{n/2}} \exp \left\{-\frac{|x|^2}{4t}
\right\}
 \ee
of variance equal to $2n$ \cite{Cra}.
Since the Shannon's entropy of
% \sout{the
% Gaussian density}
$M(\cdot ,t)$ equals
 \[
\rH (M(\cdot,t)) = \frac n2 \log (4\pi\, \rme\, t),
 \]
it follows that the \BBB corresponding \nc entropy power
% \sout{of the source-type solution to the
% heat equation}
is a linear function of time, i.e.
% \be %\label{eg}
 \begin{displaymath}
\rN (M(\cdot, t)) = 4\pi\, \rme\, t, \qquad\text{hence}\qquad
\frac{\d^2}{\d t^2}\rN (M(t)) = 0.
\end{displaymath}
% \ee
% Hence,
%   \be\label{gg}
% \frac{\d^2}{\d t^2}\rN (M(t)) = 0.
%  \ee
In view of this remark, the concavity property of entropy power can
be rephrased by saying that for all times $t >0$ the
\BBB fundamental \nc source-type solution
%\sout{is the unique one which}
% \footnote{di fatto bisognerebbe dimostrare che sulle altre la derivata
%   seconda \`e
%   strettamente negativa, immagino sar\`a vero ma non segue da quello
%   che abbiamo detto prima.}
maximizes the second derivative of the
Shannon's entropy power among all possible solutions to the
heat equation.

This idea easily extends to the nonlinear heat equation \fer{nl}.
\BBB The corresponding \nc
fundamental %\sout{example of}
solution
% \sout{to the nonlinear heat
%   equation}
was found
around 1950 by Zel'dovich and Kompaneets and Barenblatt
%\sout{, who found and
%analyzed the solution representing heat release from a point source}
\cite{Vaz}. In the case $p>1$ (see \cite{BDV} for $p <1$)
the Barenblatt (also called self-similar
or generalized Gaussian)
%\sout{source-type}
solution %(usually called Barenblatt solution or generalized Gaussian)
departing from $x=0$ takes the self-similar form
\BBB (recall the definition of $\mu$ in \eqref{eq:3})\nc
 \be\label{ba-self}
 M_p(x,t) := \frac 1{t^{\BBB n/\mu}} \tilde M_p\left (\frac x{t^{1/\mu}}\right),
   % left |
    % \frac{x}{t^{1/\mu}}}\right|^2
    %t^{-2\beta}
 \ee
 where
\begin{equation}
 \tilde M_p(x)=\big(C - \kappa\, |x|^2   \big)^{\frac
1{p-1}}_+ ;
\label{ba}
 \end{equation}
 here $(s)_+ = \max\{s, 0\}$,
$\kappa := \frac 1{2\mu}\,\frac{p-1}{p},$
and
%  \be\label{con}
%  %\alpha:= \frac n{2 + n(p-1)}, \quad
% % \beta := \frac1{\mu}=\frac 1{2 + n(p-1)},\quad
%  %\frac \alpha n, \quad
% \kappa := \frac 1{2\mu}\,\frac{p-1}{p}.
%  \ee
the constant $C$ in \fer{ba} can be chosen to fix the mass of the
source-type Barenblatt solution equal to one.

Notice that,
%In more details,
if we %given a function $g(x)\ge 0, x \in \R^n $, let us
consider \BBB the mass--preserving rescaling \nc
 \[
%\be
%\label{scal}
 \mathcal R_a:f(x) \to \mathcal R_a f(x) := {a^{-n}} f\left( x/a \right), \quad a >0,\ x\in \R^n,
\] %\ee
of a given nonnegative density $f$,
% , let us which preserves the total mass of the  function $g$.
a direct computation immediately yields
% inspection, it is immediate to conclude that the $p$-th R\'enyi
% entropy \fer{ren} is such that, if $g_a$ is defined as in \fer{scal}
 \be\label{h-scale}
 \rH_p(\mathcal R_a f) = \rH _p(f) + n \log a,\quad
 \rN_p(\mathcal R_af)=a^{\mu}\,\rN_p(f).
 \ee
\BBB An application of \eqref{h-scale} to \eqref{ba-self}
with $a:=t^{1/\mu}$ \nc
% Elementary
% computations
then shows that
%  \[
%  \rH _p(M_p(\cdot, t)) = \rH_p(\tilde M_p)+\nu^{-1} \log
%  (%C_{n,p}^{-1}\,
%  t),
%  \]
% with
%  \[
% C_{n,p} = \left( \int_{\R^n} M_p^p(x, t=1) \, \d x
% \right)^{\nu/(p-1)
%   %1 + \frac 2{n(p-1)}
% }.
%  \]
%Consequently,
the $p$-th R\'enyi entropy
power
%of R\'enyi entropy
defined in
\fer{pow-ren} is a linear function of time
\[%\be \label{eb}
\rN_p(M_p(t)) = \rN_p(\tilde M_p) \,t,
\quad
\text{so that}\quad
\frac{\d^2}{\d t^2}\rN_p(M_p(t)) = 0.
 \]%\ee
% so that
%   \be\label{gB}
% \frac{d^2}{dt^2}\rN_p(M_p(t)) = 0.
%  \ee
As before, the concavity property of R\'enyi entropy power would
imply that for all times $t >0$
the Barenblatt source-type solution maximizes
% \sout{is the unique one
% which maximizes,}
the second derivative of the
$p$-th R\'enyi entropy power.
 among all possible solutions to the nonlinear heat
equation.

\section{A new Lyapunov functional}\label{fis}

The proof of \fer{conc-p} requires to evaluate two time derivatives
of the $p$-th R\'enyi entropy power, along the solution to the
nonlinear heat equation \fer{nl}. The first derivative of the
entropy power can be  easily obtained: in fact, if $u(\cdot,t)$ solves
the nonlinear heat equation \fer{nl}, integration by
parts immediately leads to (cfr. Appendix \ref{app:1})
 \be\label{e-r}
 \frac \d{\d t} \rH _p(u(\cdot,t)) =  \rI_p(u(\cdot,t)), \quad t >0,
 \ee
where
 \be\label{fis-r}
 \cI_p(X)= \rI_p(f) := \frac 1{\int_{\R^n} f^p \, \d x}
   \int_{\{f>0\}} \frac{|\nabla f^p(x)|^2}{f(x)} \, \d x.
 \ee
When $p \to 1$, identity \fer{e-r} reduces to DeBruijn's identity, which
connects Shannon's entropy functional with the Fisher information of a
random variable with density
 \be\label{fish}
 \cI(X)= \rI(f) := \int_{\R^n} \frac{|\nabla f(x)|^2}{f(x)} \, \d x,
 \ee
via the heat equation.
% \sout{Indeed, if $u(v,t)$ denotes the solution to the
% heat equation \fer{heat}, integration by parts immediately leads to the
% relationship \fer{e-r}.}
%\footnote{L'abbiamo gi\`a detto per $p\neq 1$}
 Analogous computations can be done here.  Using
identity \fer{e-r} we get
 \[
\frac \d{\d t} \rN_p(u(\cdot, t)) =
  \nu\,
  %\left(\frac 2n +p-1\right)
  \rN_p(u(\cdot,t)) \, \rI_p(u(\cdot,t)).
 \]
% \sout{For a given probability density $f$, let $u(x,t)$ denote the
% solution to the nonlinear heat equation \fer{nl} with initial value
% $u(x, t=0) = f(x)$, and}
Let us introduce the quantity
 \be\label{Costa}
\BBB
\Upsilon_p(u) := \rN_p(u)\, \rI_p(u),\quad
\upsilon(t):=\Upsilon_p(u(\cdot,t)).\nc
  \ee
Since $2/n+p-1>0$, the concavity of entropy power can be rephrased
as the decreasing in time property of
$t\mapsto \Upsilon_p(u(\cdot,t))$.
% For the rest of this Section. let us assume that
% the concavity of R\'enyi entropy power has been proven.

It is important to notice that
\BBB by \eqref{h-scale} and
the scaling property of the $p$-th Fisher information
\be\label{f-scale}
\rI_p(\mathcal R_a f) = a^{-\mu %2 + n(p-1)
} \,\rI_p(f),
 \ee \nc
% our choice of R\'enyi entropy
% power,
the functional $\Upsilon_p$ is invariant under
the dilations $\mathcal R_\cdot$,\ i.e.
% In more details,
% if we %given a function $g(x)\ge 0, x \in \R^n $, let us
% consider \BBB the mass--preserving rescaling \nc
%  \be\label{scal}
%  f(x) \to f_a(x) := {a^n} f\left( {a}x \right), \quad a >0,\ x\in \R^n,
%  \ee
% of a given nonnegative density $f$,
% , let us which preserves the total mass of the  function $g$.
%a direct computation immediately yields
% inspection, it is immediate to conclude that the $p$-th R\'enyi
% entropy \fer{ren} is such that, if $g_a$ is defined as in \fer{scal}
 % \be\label{h-scale}
 % \rH_p(f_a) = \rH _p(f) - n \log a,\quad
 % \rN_p(f_a)=a^{-\mu}\,\rN_p(f_a).
 % \ee
% Since the $p$-th Fisher's information \fer{fis-r} scales according to

% one concludes that the functional $\Upsilon_p(f)$ is invariant respect to
% the scaling \fer{scal}, i.e.
 \be\label{np}
 \Upsilon_p(\mathcal R_a f) =\Upsilon_p(f)\quad
 \text{for every }a>0.
 \ee
Property \fer{np} allows to identify the long-time behavior of the
function $\upsilon (t)=\Upsilon_p(u(\cdot,t))$.
% \sout{Unless the initial density $f(x)$ in the
% nonlinear heat equation is a Dirac delta function (which generates
% the source-type Barenblatt solution),}
% \footnote{noi non sappiamo
% caratterizzare
% veramente il caso lineare}
%the functional $\Upsilon_p_f(t)$
It is nonincreasing, and it will reach its \BBB infimum
\nc
%\sout{eventual minimum value}
as time $t\to \infty$. The computation of the limit value uses in a
substantial way the scaling invariance property. In fact,
\BBB we can rescale \nc
$u(x,t)$ according to
\begin{equation}\label{FP}
 U(x,t) =
 \BBB t^{-n/\mu}\, u(x\, t^{-1/\mu}, t)=
 \mathcal R_{t^{1/\mu}}u(\cdot,t)
\end{equation}
\BBB where $\mu$ is defined in \eqref{eq:3} as usual, so that
\begin{displaymath}
  \upsilon(t)=\Upsilon_p(u(\cdot,t))=\Upsilon_p(U(\cdot,t))
  \quad\text{for every }t>0.
\end{displaymath}
\nc
%$\alpha$ and $\beta$ are defined in \fer{con}.
% at each
%time $t>0$, the value of $\Upsilon_p_f(t)$ does not change if On the other
On the other hand, it is well-known that (cfr. for example \cite{Vaz})
 \be\label{limi}
 \lim_{t\to \infty} U(x,t) = %\tilde
 \tilde M_p(x),
 \ee
% \sout{where $\tilde M_p(x)$ is}
the Barenblatt profile defined in
\fer{ba}.
Therefore, \BBB denoting by $f(x):=u(x,0)$ the initial
probability density, \nc
passing to the limit one obtains
% , for any given
% probability density $f$,
the (isoperimetric) inequality for the $p$-th R\'enyi entropy:
\begin{thm}
  If $p > n/(n+2)$
  every smooth, strictly positive and rapidly decaying probability
  density $f$
  satisfies
  \be\label{b5} \Upsilon_p (f)=\rN_p(f) \, \rI_p(f) \ge \rN_p(\tilde
  M_p) \, \rI_p(\tilde M_p) = :
  % \Upsilon_p(\tilde M_p)=:
  \gamma_{n,p}, \ee
  where the value of the strictly positive
  constant $\gamma_{n,p}$ is
  given \eqref{val1} and \eqref{val2} of
  Appendix \ref{const}.
\end{thm}
\nc

\section{Sobolev inequality revisited}
Inequality \fer{b5} can be rewritten in a form more suitable to functional
analysis. Let $f(x)$ be a probability density in $\R^n$. Then, if $p >
n/(n+2)$
 \be\label{gn}
\int_{\R^n} \frac{|\nabla f^p(x)|^2}{f(x)} \, \d x \ge \gamma_{n,p} \left(
\int_{\R^n} f^p(x) \, \d x \right)^{\frac{ 2+\BBB 2n \nc (p-1)}{n(p-1)}}.
 \ee
If $n >2$, the case $p= (n-1)/n$ is distinguished from the others, since
it leads to
 \[
\frac{2+\BBB 2n \nc (p-1)}{n(p-1)}= 0,\quad  \nu=\frac 1n,
 \]
 and
\begin{displaymath}
  \rN_{1-1/n}(f)=\int_{\R^n}f^{1-1/n}(x)\,\d x.
\end{displaymath}
In this case the concavity of $\rN_{1-1/n}$ along
\eqref{nl}
has been already known and has a nice geometric interpretation
in terms of transport distances, see \cite{Otto01}. \nc

Note that the restriction $n>2$ implies $(n-1)/n > n/(n+2)$.
Hence, for $p= (n-1)/n$ we obtain that the probability density $f$
satisfies the inequality
 \be\label{sob12}
\int_{\R^n} \frac{|\nabla f^{(n-1)/n}(x)|^2}{f(x)} \, \d x \ge
\gamma_{n,(n-1)/n}.
%n\pi\,
%\frac{2^2(n-1)^2}{n-2}\left(\frac{\Gamma\left(
% n/2 \right)}{\Gamma\left(n \right)} \right)^{2/n}.
 \ee
The substitution
%Let us observe that, if we set
$f= g^{2^*}$, \BBB where $2^*=2n/(n-2)$, yields
\nc
% the choice
%  \[
%  \mu = \frac 2{2p-1}
%  \]
%is such that
 \[
 \int_{\R^n} \frac{|\nabla f^{(n-1)/n}(x)|^2}{f(x)} \, \d x =
%\left( \frac{2p}{2p-1}\right)^2
 \left(\frac{2n-2}{n-2}\right)^2\int_{\R^n} |\nabla g(x)|^2 \,
 \d x.
 \]
Therefore,
%if $p = (n-1)/n$,
for any given function $g \ge 0$ such that
$g(x)^{2^*}$ is a probability density in $\R^n$, with $n>2$, we
obtain the inequality
 \be\label{sob14}
\int_{\R^n} |\nabla g(x)|^2 \,
 \d x \ge
\left(\frac {n-2}{2n-2}\right)^2\,\gamma_{n,(n-1)/n}.
% n(n-2) \pi \, \left(\frac{\Gamma\left(
%  n/2 \right)}{\Gamma\left(n \right)} \right)^{2/n}.
 \ee
A \BBB careful \nc
%\sout{simple}
computation \BBB (see Appendix A) \nc gives
 \[
\gamma_{n,(n-1)/n} = n\pi\, \frac{2^2(n-1)^2}{n-2}\left(\frac{\Gamma\left(
 n/2 \right)}{\Gamma\left(n \right)} \right)^{2/n},
 \]
% Hence, for $p= (n-1)/n$ we obtain that the probability density $f$
% satisfies the inequality
%  \be\label{sob12}
% \int_{\R^n} \frac{|\nabla f^{(n-1)/n}(x)|^2}{f(x)} \, \d x \ge n\pi\,
% \frac{2^2(n-1)^2}{n-2}\left(\frac{\Gamma\left(
%  n/2 \right)}{\Gamma\left(n \right)} \right)^{2/n}.
%  \ee
%
and simple scaling argument finally shows that, if $g(x)^{2^*}$ has a
mass different from $1$, $g$ satisfies the \emph{Sobolev} inequality
\cite{Aub}, \cite{Tal}
 \be\label{sob15}
 \int_{\R^n} |\nabla g(x)|^2 \,
 \d x
\ge \mathcal{S}_n \left(\int_{\R^n} g(x)^{2^* %n/(n-2)
  }\,\d x \right)^{2/2^*},%(n-2)/n},
 \ee
where
 \[
\mathcal{S}_n  = n(n-2) \pi \, \left(\frac{\Gamma\left(
 n/2 \right)}{\Gamma\left(n \right)} \right)^{2/n}
 \]
is the sharp Sobolev constant. Hence, Sobolev inequality with the sharp
constant is a consequence of the concavity of  R\'enyi entropy power of
parameter $ p = (n-1)/n$, when $n >2$.

In all the other cases, the concavity of R\'enyi entropy power leads to
Gagliardo-Nirenberg type inequalities with sharp constants, like the ones
recently studied by Del Pino and Dolbeault \cite{DD}, and
Cordero-Erausquin, Nazaret, and Villani, \cite{CNV} with different
methods.

\section{Proof of the concavity of R\'enyi entropy
power}\label{sec:proof}

Arguing as in \cite{Vaz2}, it is sufficient to consider the case
of smooth, strictly positive and rapidly decaying probability densities.
% Let us introduce some notation.

\nc
For a given probability
density $u$ we set
 \[
\rE_p(u):=%\frac1{p-1}\int u^p\, \d x=
\int \sfep(u(x))\,\d x,
\quad \text{where }
\sfep(r):=\frac 1{p-1}\,r^p  .
 \]
% where
%  \[
%  \sfep(r):=\frac 1{p-1}\,r^p  .
%  \]
 Consequently, the $p$-th R\'enyi entropy of $u$ can be written as
 \[
  \rH _p(u):=
\frac1{1-p}\log\Big((p-1)\rE_p(u)\Big).
 \]
Likewise, since
 $
\sfep'(u)=\sffp\,u^{p-1},\ \sffp:= \frac p{p-1},
 $
  we can write
 \begin{align*}
   \rF_p(u):=&\int \frac{|\rmD\, u^p|^2}{u}\,\d x= \sffp^2 \int |\rmD\,
   u^{p-1}|^2\,u\,\d x\\=&
    \int
\big|\rmD\sfep'(u)\big|^2 u\,\d x.
 \end{align*}
 Recall also that
 \[
L_p(u)=u^p=u\,\sfep'(u)-\sfep(u),\quad L_p'(u)=u\,\sfep''(u).
 \]
The nonlinear heat equation \fer{nl} can be equivalently written as
 \be\label{nl2}
\partial_t u-\nabla\cdot\Big(u \,\rmD \sfep'(u)\Big)=0.
 \ee
Using equation \fer{nl2}, and integrating by parts, we obtain
\BBB (see Appendix B) \nc
\begin{equation}
    \label{eq:1}
    -\frac{\d}{\d t}\rE_p(u_t)=\rF_p(u),
\end{equation}
 \begin{align}
   \label{eq:2}
    \rD_p(u_t):=&-\frac \d{\d t}\rF_p(u_t)\\=&\notag
     2\int u^p \Big(|\rmD^2 \sfep'(u)|^2
    +(p-1) \big(\Delta\sfep'(u)\big)^2\Big)\,\d x.
\end{align}
For a given function $\phi_t$ which depends on time, and a positive
constant $\sigma$
\begin{displaymath}
    \frac{\d^2}{\d t^2}\exp(\sigma\, \phi_t)=
    \exp(\sigma\phi_t)\Big(\sigma\phi_t''+(\sigma\phi_t')^2\Big).
\end{displaymath}
Therefore
 \[
    \frac{\d^2}{\d t^2}\exp(\sigma\phi_t)\le 0
    \qquad\Longleftrightarrow\qquad
    -\sigma\phi_t''\ge (\sigma\phi_t')^2.
 \]
 % if and only if
 % \[
 %    -\sigma\phi_t''\ge (\sigma\phi_t')^2.
 % \]
If $\phi_t:=\rH _p(u_t)$, where $u_t=u(\cdot,t)$, we have
\begin{equation}
    \label{eq:4}
    \phi_t'=\frac 1{p-1}\frac{-\rF_p(u_t)}{\rE_p(u_t)},
 \ee
 and
 \be\label{eq:5}
 \phi_t''=\frac
    1{p-1}\frac{\rD_p(u_t)\,\rE_p(u_t)-\rF_p(u_t)^2}{\rE_p(u(t))^2}.
\end{equation}
Hence we end up with the condition
\begin{displaymath}
 \frac \sigma{p-1}\Big(\rD_p(u_t)\,\rE_p(u_t) - \rF_p(u_t)^2\Big)\ge
    \Big(\frac \sigma{p-1}\Big)^2\rF_p(u_t)^2,
\end{displaymath}
i.e.~(suppressing the index $t$)\nc
\begin{displaymath}
%    \label{eq:6}
    \sigma \rD_p(u)\int u^p\,\d x\ge
    \big(\sigma^2 + \sigma(p-1)\big)\rF_p(u)^2.
\end{displaymath}
Since $\sigma>0$, the second derivative is non positive if
\begin{equation}
    \label{eq:7}
    \rD_p(u)\int u^p\,\d x\ge \big(\sigma + (p-1)\big)\big(\rF_p(u)\big)^2.
\end{equation}
Since
\begin{displaymath}
  \rF_p(u)=\int \rmD u^p\cdot \rmD \sfep'(u)\,\d x=
    -\int u^p\,\Delta \sfep'(u)\,\d x,
\end{displaymath}
by Cauchy-Schwarz inequality we have
\begin{displaymath}
  \rF_p(u)^2\le \int u^p\,\d x\,\int u^p(\Delta \sfep'(u))^2\,\d x.
\end{displaymath}
It follows that \eqref{eq:7} holds if
\begin{displaymath}
    \rD_p(u)\ge (\sigma + (p-1))\int u^p (\Delta \sfep'(u))^2\,\d x.
\end{displaymath}
On the other hand, by the trace inequality,
$|\rmD^2 f|^2\ge \frac 1n \big(\Delta f\big)^2$
\begin{displaymath}
  \rD_p(u)\ge 2\,(\frac 1n+(p-1))\, \int u^p (\Delta \sfep'(u))^2\,\d x,
\end{displaymath}
and we end up with the condition
\begin{equation}
    \label{eq:8}
    \sigma\le \frac 2n+ p-1=\nu.
\end{equation}
Choosing for $\sigma$ the upper bound in \fer{eq:8} we conclude.

\appendices

\section{Computation of the constants}\label{const}

Let us recall here some useful formulas. The surface of the $n-1$
dimensional unit sphere $\smooth^{n-1}$ is given by $|\smooth^{n-1}| =
2\pi^{n/2}/ \Gamma(n/2)$. Let us first consider the case $p>1$. If $a >0$,
using the integral representation of  Beta function we have
 \begin{align*}
 &\int_{\R^n} (1-|x|^2)_+^a \, \d x = |\smooth^{n-1}|\int_0^1
 \rho^{n-1}(1-\rho^2)^a \, d\rho =
\\&\frac{2\pi^{n/2}}{\Gamma(n/2)}\int_0^1
 t^{n/2-1}(1- t)^a \, dt = \frac{2\pi^{n/2}}{\Gamma(n/2)} B(\frac n2, a+1)=\\&
 \pi^{n/2}\frac{\Gamma(a+1)}{\Gamma\left(\frac n2 + a+1\right)}.
\end{align*}
With this formula, we can evaluate quantities associated to the Barenblatt
function \fer{ba}. Indeed, if
 \be\label{m1}
A_p = \int_{\R^n} (1-|x|^2)_+^{1/(p-1)} \, \d x =
\pi^{n/2}\frac{\Gamma\left(\frac{p+1}p\right)}{\Gamma\left(\frac n2
+\frac{p+1}p \right)},
 \ee
we obtain a Barenblatt of mass equal to one, we denote by
 \[
\mathcal{B}_p(x) = (C_p-|x|^2)_+^{1/(p-1)}
 \]
if
 \be\label{con1}
C_p = A_p^{-\frac{2(p-1)}{n(p-1)+2}}.
 \ee
Also \cite{Vaz}
 \be\label{mom}
\int_{\R^n} |x|^2 \mathcal{B}_p(x)\, \d x = \frac{n(p-1)}{(n+2)p -n} C_p,
 \ee
and, since
 \[
\int_{\R^n} \mathcal{B}_p(x)^p \, \d x = \int_{\R^n}(C_p- |x|^2)
\mathcal{B}_p(x)\, \d x,
 \]
one obtains
 \be\label{po}
\int_{\R^n} \mathcal{B}_p(x)^p \, \d x = \frac{2p}{(n+2)p -n} C_p.
  \ee
Thanks to \fer{mom} and \fer{po}, we reckon the values of the $p$-Fisher
information $\rI_p(f)$ defined in \fer{fis-r} and of the R\'enyi entropy
$\rH _p(f)$, associated to $\mathcal{B}_p$
 \be\label{rb}
\rH _p(\mathcal{B}_p) = \frac 1{1-p} \log \frac{2p}{(n+2)p -n} C_p,
 \ee
and
 \be\label{fb}
\rI_p(\mathcal{B}_p) = n\frac{2p}{p-1}.
 \ee
Hence, if $p>1$ the value of the constant $\gamma_{n,p}$ is
 \begin{align}\label{val1}
&\gamma_{n,p} =  n\pi\frac{2p}{p-1}\cdot \nonumber \\& \cdot\left(
\frac{\Gamma\left(\frac{p+1}p\right)}{\Gamma\left(\frac n2 +\frac{p+1}p
\right)}\right)^{2/n}\left(\frac{(n+2)p -n}{2p} \right)^{\frac{2+
n(p-1)}{n(p-1)}}.
 \end{align}

Analogous computations can be done in the case $p<1$. In this case
 \be\label{m2}
A_p = \int_{\R^n} (1+|x|^2)^{1/(p-1)} \, \d x = \pi^{n/2}\frac{\Gamma\left(
\frac 1{1-p} - \frac n2  \right)}{\Gamma\left(\frac 1{1-p}\right)},
 \ee
while
 \be\label{mom2}
\int_{\R^n} |x|^2 \mathcal{B}_p(x)\, \d x = \frac{n(1-p)}{(n+2)p -n} C_p.
 \ee
Note that, if $p<1$, the second moment of the Barenblatt is bounded if and
only if $p > n/(n+2)$. Therefore, the computations that follow are
restricted to this domain of $p$. If this is the case, formula \fer{po}
still holds, while
 \be\label{fb2}
\rI_p(\mathcal{B}_p) = n\frac{2p}{1-p}.
 \ee
Finally, if $n/(n+2) < p <1$,
 \begin{align}\label{val2}
&\gamma_{n,p} =  n\pi\frac{2p}{1-p}\cdot \nonumber \\&
\cdot\left(\frac{\Gamma\left( \frac 1{1-p} -\frac n2
\right)}{\Gamma\left(\frac 1{1-p}\right)} \right)^{2/n}\left(\frac{(n+2)p
-n}{2p} \right)^{\frac{2+ n(p-1)}{n(p-1)}}.
 \end{align}

\section{Proof of identities \eqref{eq:1} and \eqref{eq:2}}
\label{app:1}

\noindent\eqref{eq:1} follows by a simple
integration by parts:
\begin{align*}
    \frac{\d}{\d t}\rE_p(u_t)&=\int \sfep'(u)\,\partial_t u\,\d x=
    \int \sfep'(u)\,\nabla\cdot\Big(u \,\rmD \sfep'(u)\Big)\,\d x
    \\&=
    -\int \rmD\sfep'(u)\cdot \rmD \sfep'(u)\, u\,\d x.
\end{align*}
\eqref{eq:2} is based on the Bochner identity or, equivalently,
Bakry-\'Emery $\Gamma$-calculus
(in their simplest Euclidean form),
see \cite{OW} for analogous computations of
the second derivative of $\rE_p$ along geodesics in the Wasserstein space.
 \nc

One has
\begin{align*}
    \partial_t \sfep'(u)&=\sfep''(u)\partial_t u=
%   \sfep''(u)\,\Delta u^p=
    \sfep''(u)\,\nabla\cdot\Big(u \,\rmD \sfep'(u)\Big)
    \\&=
    u\,\sfep''(u)\,\Delta \sfep'(u)+
    \sfep''(u)\, \rmD u\cdot\rmD \sfep'(u)
    \\&=L_p'(u)\,\Delta \sfep'(u)+ \big|\rmD\sfep'(u)\big|^2.
\end{align*}
Recall the Bochner identity
\begin{align*}
    2\Gamma_2(g)=\Delta |\rmD g|^2-2\rmD g\cdot \rmD\Delta g=
    2|\rmD^2 g|^2,
\end{align*}
and
 \[
    (rL_p'(r)-L)'=rL_p''(r),
    \]
while
    \[
    (rL_p'(r)-L)'\sfep''(r)=L_p''(r)L_p'(r).
 \]
Then
\begin{align*}
    &\frac \d{\d t}\rF_p(u_t)
    =  \int \Big(\big|\rmD \sfep'(u)\big|^2 \partial_t u +
    2u\,\rmD\sfep'(u)\cdot \rmD \partial_t \sfep'(u)\Big)\,\d x
    \\&=
    \int \big|\rmD \sfep'(u)\big|^2 \Delta u^p\,\d x
    \\&\quad + 2\int
    u\,\rmD\sfep'(u)\cdot \rmD\Big(L_p'(u)\,\Delta \sfep'(u)+ \big|\rmD\sfep'(u)\big|^2\Big)\,\d
    x
    \\&=
    \int u^p \Delta\big|\rmD \sfep'(u)\big|^2\,\d x+
    2\int uL_p'(u)\,\rmD\sfep'(u)\cdot \rmD \Delta \sfep'(u)\,\d x
    \\&\quad+
    2\int \rmD L_p(u)\cdot \rmD L_p'(u)\,\Delta\sfep'(u)\,\d x
    -2 \int L_p(u) \Delta \big|\rmD\sfep'(u)\big|^2
    \\&=
    -\int L_p(u) \Delta\big|\rmD \sfep'(u)\big|^2\,\d x+
    2\int L_p(u)\,\rmD\sfep'(u)\cdot \rmD \Delta \sfep'(u)\,\d x
    \\&\quad+ 2\int \big(uL_p'(u)-L_p(u)\big)\,\rmD\sfep'(u)\cdot \rmD \Delta \sfep'(u)\,\d
    x
    \\&\quad
    +2\int \rmD L_p(u)\cdot \rmD L_p'(u)\,\Delta\sfep'(u)\,\d x
    \\&=
    -\int L_p(u) \Gamma_2(\sfep'(u))\,\d x\\&\quad
    -2\int \big(uL_p'(u)-L_p(u)\big)\,\big(\Delta\sfep'(u)\big)^2\,\d x
    \\&=
    -2\int L_p(u) |\rmD^2 \sfep'(u)|^2
        +\big(uL_p'(u)-L_p(u)\big)\,\big(\Delta\sfep'(u)\big)^2\,\d
        x
    \\&=
    -2\int u^p \Big(|\rmD^2 \sfep'(u)|^2
    +(p-1) \big(\Delta\sfep'(u)\big)^2\Big)\,\d x.
\end{align*}


\ifCLASSOPTIONcaptionsoff
  \newpage
\fi

\end{document}